\documentclass[journal]{IEEEtran}

\usepackage[compress]{cite}
\usepackage{amsthm, amsmath, stackrel}
\usepackage{amssymb}
\usepackage{graphicx}
\usepackage{mathrsfs}
\usepackage{mathtools}
\usepackage{savesym}
\usepackage{cuted}
\usepackage{nccmath}
\usepackage{amsfonts}
\usepackage{epstopdf}
\usepackage{lipsum}
\usepackage{float}
\usepackage{stfloats}
\usepackage{url}
\usepackage{subcaption}
\usepackage{algorithm}
\usepackage{algpseudocode}
\usepackage{algcompatible}
\usepackage[super]{nth}
\usepackage[english]{babel}
\usepackage{multirow}
\usepackage{makecell}
\usepackage{soul}
\usepackage[dvipsnames]{xcolor}

\providecommand{\U}[1]{\protect\rule{.1in}{.1in}}

\makeatother

\begin{document}

\title{Exploration and Exploitation in Federated Learning to Exclude Clients with Poisoned Data}

\author{Shadha Tabatabai\IEEEauthorrefmark{1}\IEEEauthorrefmark{2},
Ihab Mohammed\IEEEauthorrefmark{1},
Basheer~Qolomany\IEEEauthorrefmark{3}, Abdullatif Albasser\IEEEauthorrefmark{4}, Kashif Ahmad\IEEEauthorrefmark{4}, \\ Mohamed Abdallah\IEEEauthorrefmark{4}, Ala Al-Fuqaha\IEEEauthorrefmark{4} \\
\IEEEauthorblockA{\IEEEauthorrefmark{1} School of Computer Sciences, Indiana Institute of Technology, USA. \\
\{smtabatabai,iamohammed@indianatech.edu\}} \\
\IEEEauthorblockA{\IEEEauthorrefmark{2} Department of Computer Science, Western Michigan University, USA, shadhamuhinoo.tabatabai@wmich.edu} \\
\IEEEauthorblockA{\IEEEauthorrefmark{3} Department of Cyber Systems, College of Business \& Technology, University of Nebraska at Kearney,  Kearney, NE 68849, qolomanyb@unk.edu} \\
\IEEEauthorblockA{\IEEEauthorrefmark{4} Information and Computing Technologies (ICT) Division, College of Science and Engineering (CSE), Hamad Bin Khalifa University, Doha, Qatar.\\ \{amalbaseer,kahmad,moabdallah,aalfuqaha@hbku.edu.qa\}}
}

\maketitle

\begin{abstract}

Federated Learning (FL) is one of the hot research topics, and it utilizes Machine Learning (ML) in a distributed manner without directly accessing private data on clients. However, FL faces many challenges, including the difficulty to obtain high accuracy, high communication cost between clients and the server, and security attacks related to adversarial ML. To tackle these three challenges, we propose an FL algorithm inspired by evolutionary techniques. The proposed algorithm groups clients randomly in many clusters, each with a model selected randomly to explore the performance of different models. The clusters are then trained in a repetitive process where the worst performing cluster is removed in each iteration until one cluster remains. In each iteration, some clients are expelled from clusters either due to using poisoned data or low performance. The surviving clients are exploited in the next iteration. The remaining cluster with surviving clients is then used for training the best FL model (i.e., remaining FL model). Communication cost is reduced since fewer clients are used in the final training of the FL model. To evaluate the performance of the proposed algorithm, we conduct a number of experiments using FEMNIST dataset and compare the result against the random FL algorithm. The experimental results show that the proposed algorithm outperforms the baseline algorithm in terms of accuracy, communication cost, and security.

\end{abstract}

\IEEEpeerreviewmaketitle

\section{Introduction}
\label{introduction}


Recently Federated learning (FL) has been proposed as an emerging approach to build Machine Learning (ML) models across multiple decentralized edge devices~\cite{albaseer2021fine}. This helps to overcome the challenge of privacy preservation by keeping all the training data on the device, decoupling the ability to do ML from the need to store the data in the cloud. 
However, several challenges need to be considered for the implementation of FL including communication cost between the servers and clients, the accuracy of the model, and security.


FL is vulnerable to security attacks whereby a group of malicious clients could harm the performance of the model by carrying out a poisoning attack \cite{mothukuri_survey_2021}. These attacks may cause the model to fail and converge to biased models that do not accurately represent the data. Applying anti-poisoning techniques might lead to the discrimination of minority groups whose data are significantly and legitimately different from those of the majority of clients \cite{singh_fair_2020}. In addition, detection and identification of unauthorized IoT devices are very important especially with the increase in the number of attacks on IoT devices.

In this paper, to cope with the FL challenges, we propose an FL framework inspired by evolutionary techniques consisting of three stages. In the first stage, participating clients are grouped randomly into a number of clusters. Subsequently, a random model is selected for each cluster. In the second stage, models are explored and the best performing cluster, in terms of classification accuracy, is selected in a repetitive process. In each iteration, all the clusters are trained in parallel and the worst performing cluster is removed from the process till one cluster remains. Additionally, in each iteration, several clients may be expelled from each cluster either due to their low performance compared with other clients in the cluster or their data being poisoned. The remaining clients of removed clusters are exploited by joining the best-performing cluster. In the third stage, the best performing cluster is utilized such that FL is trained with the cluster's model using the remaining clients in that cluster.

The salient \textit{contributions of this paper} are:
\begin{itemize}

\item Optimize the performance of FL in terms of accuracy by exploring a number of clusters, each with a different model, to select the best performing model (i.e., cluster).

\item Optimize the security of FL by identifying clients with poisoned data and expel them from every cluster using cosine similarity while surviving clients are exploited in every iteration.

\item Optimize the communication cost during the training process by expelling clients with either poisoned or weak data. Thus fewer clients participate in the training process leading to less communication.
\end{itemize}

The organization of the remainder of the paper is as follows. Related literature is reviewed in Section~\ref{related_work}. The system model is described in Section~\ref{system_model}. Section~\ref{proposed_algorithm} discusses the proposed algorithm. The used dataset and conducted experiments are explained in Section~\ref{experimental_settings}. A discussion of the results and the salient lessons learned are provided in Section~\ref{result_discussion}. Finally, the paper is concluded in Section~\ref{conclusion_future_work} by summarizing the work and identifying future research directions.

\begin{table}[]
 \centering
 \caption{An overview of the related work.}
 \label{ta:related_work}
\begin{tabular}{|c|c|c|c|c|}
\hline
\multirow{2}{*}{\textbf{Ref.}} & \multirow{2}{*}{\textbf{Year}} & \multicolumn{3}{c|}{\textbf{Target/Focus}} \\ \cline{3-5} & & Accuracy  & Communication cost  & Security \\ \hline
\cite{mohammed_budgeted_2020} &  2020     &  \checkmark    &       & 
\\ \hline
\cite{ahmed2020active} &   2020  & \checkmark     &   &     
\\ \hline
\cite{qolomany_particle_2020} &   2020  &      & \checkmark  &
\\ \hline
\cite{yao_towards_2019}  &  2019   &      & \checkmark  &   
\\ \hline
\cite{sattler_clustered_2020} & 2020  &  \checkmark   &  &  
\\ \hline
\cite{singh_fair_2020} & 2020  &  &  & \checkmark
\\ \hline
\cite{doku_mitigating_2021} &  2021 & &  & \checkmark
\\ \hline
\cite{zhang_poisongan_2021} &  2021 &     &  &   \checkmark
\\ \hline
\cite{liu_secure_2020} &  2020 & &  &  \checkmark
\\ \hline
\textbf{This Work} &  2021 & \checkmark & \checkmark &  \checkmark \\ \hline
 
\end{tabular}
\end{table}

\section{Related Work}
\label{related_work}

Recently, a significant amount of work has been done in the area of FL. This section reviews recent related works on the different aspects of the work, including algorithm optimization and poisoning attacks against FL. The reviewed papers focus on optimizing the algorithm used in FL to gain more accuracy, reduce communication between the clients and the server, or enhance security. To the best of our knowledge, this work is the first attempt that optimizes FL model accuracy, reduces the number of client-server communication rounds, and enhances the security of FL models. We describe and compare the most relevant previous works with our work in Table \ref{ta:related_work}.

\subsection{Algorithm Optimization}

Several interesting optimization techniques have been proposed to deal with the challenges associated with FL, including learning an ML model in an FL environment, unbalanced distribution of local data, and reduce the generated traffic in the network.

Mohammed \emph{et al.} \cite{mohammed_budgeted_2020} proposed a stateful FL heuristic algorithm to solve the problem of optimizing accuracy in stateful FL with a budgeted number of candidate clients by selecting the best candidate clients in terms of test accuracy to participate in the training process. Ahmed \emph{et al.} \cite{ahmed2020active} tried to improve the accuracy of the FL model by employing unlabeled data available at each client through an active learning scheme. Qolomany \emph{et al.} \cite{qolomany_particle_2020} proposed a Particle Swarm Optimization (PSO)-based technique to optimize the hyperparameter settings for the local ML models in an FL environment. They evaluated and compared the proposed approach with the grid search technique. They found that the number of communication rounds used by their proposed approach is two orders of magnitude less than the grid search method. To address the issue of local clients’ data distributions diverge, Sattler \emph{et al.} \cite{sattler_clustered_2020} proposed clustered multitask FL framework, which exploits geometric properties of the FL loss surface to group the client population into clusters with jointly trainable data distributions. They found that the cosine similarity between the weight-updates of different clients is highly indicative of the similarity of their data distributions. Yao \emph{et al.} \cite{yao_towards_2019} proposed a feature fusion method by aggregating the features from both the local and global models to address the problem of high communication round cost when the local data is distributed in a Non-IID way. Yao and Sun \cite{yao_continual_2020} proposed a local continual training strategy to address the problem of weight divergence of ML model in FL environment by evaluating the important weight matrix on a small proxy dataset on the central server and then used to constrain the local training.

\subsection{Poisoning Attacks Against Federated Learning}

The communication protocol amongst different nodes in the FL environment could be exploited by attackers to launch data poisoning attacks, which has been demonstrated as a big threat to most ML models. To improve the robustness of real-world ML systems, it is critical to study how well these models perform under poisoning attacks.

To this aim, Singh \emph{et al.} \cite{singh_fair_2020} proposed two approaches to distinguish malicious behaviors of a node from legitimate ones in FL. The first approach is based on micro aggregation, with this approach, clients who identify themselves as belonging to a minority group announce some relevant attributes to their peers, such as gender, sexual orientation, or their ethnicity. While the second approach is based on Gaussian mixture models to characterize the distribution of the client-provided updates. Doku and Rawat \cite{doku_mitigating_2021} proposed an approach based on an SVM model for data vetting process to mitigate data poisoning attacks in an FL setting. They introduced the concept of a facilitator that gets assigned to an end device. The facilitator’s job is to ensure the data that an end device owns has not been compromised. Zhang \emph{et al.} \cite{zhang_poisongan_2021} proposed a poisoning attack model based on generative adversarial networks to explore an active and powerful attack model, poisoning attacks, in FL-aided IoT systems. They designed a poison data generation method to eliminate the conventional attacking assumption that the attacker already owns a proportion of other participants’ training data. Cao \emph{et al.} \cite{cao_understanding_2019} proposed a scheme, Sniper, to eliminate poisoned local models from malicious participants during training. Sniper identifies benign local models by solving a maximum clique problem, and poisoned local models will be ignored during global model updating. They analyzed how the number of poisoned samples and the number of attackers as variables affecting the performance of distributed poisoning attacks. They observed that the attack success rate increases linearly with the number of poisoned samples. The attack success rate increases with the number of attackers when the number of poisoned samples is unchanged and the increasing speed becomes faster when more attackers are involved. Liu \emph{et al.} \cite{liu_secure_2020} proposed a blockchain-based secure FL framework to address data privacy leakage issues related to ensuring secure FL in 5G networks. They used smart contracts in blockchain to validate the model updates against poisoning attacks automatically, they also introduced the local differential privacy technique in smart contracts to prevent membership inference attacks.


\section{System Model}
\label{system_model}

\begin{figure}[htbp]
\centering
\includegraphics[width=0.49\textwidth]{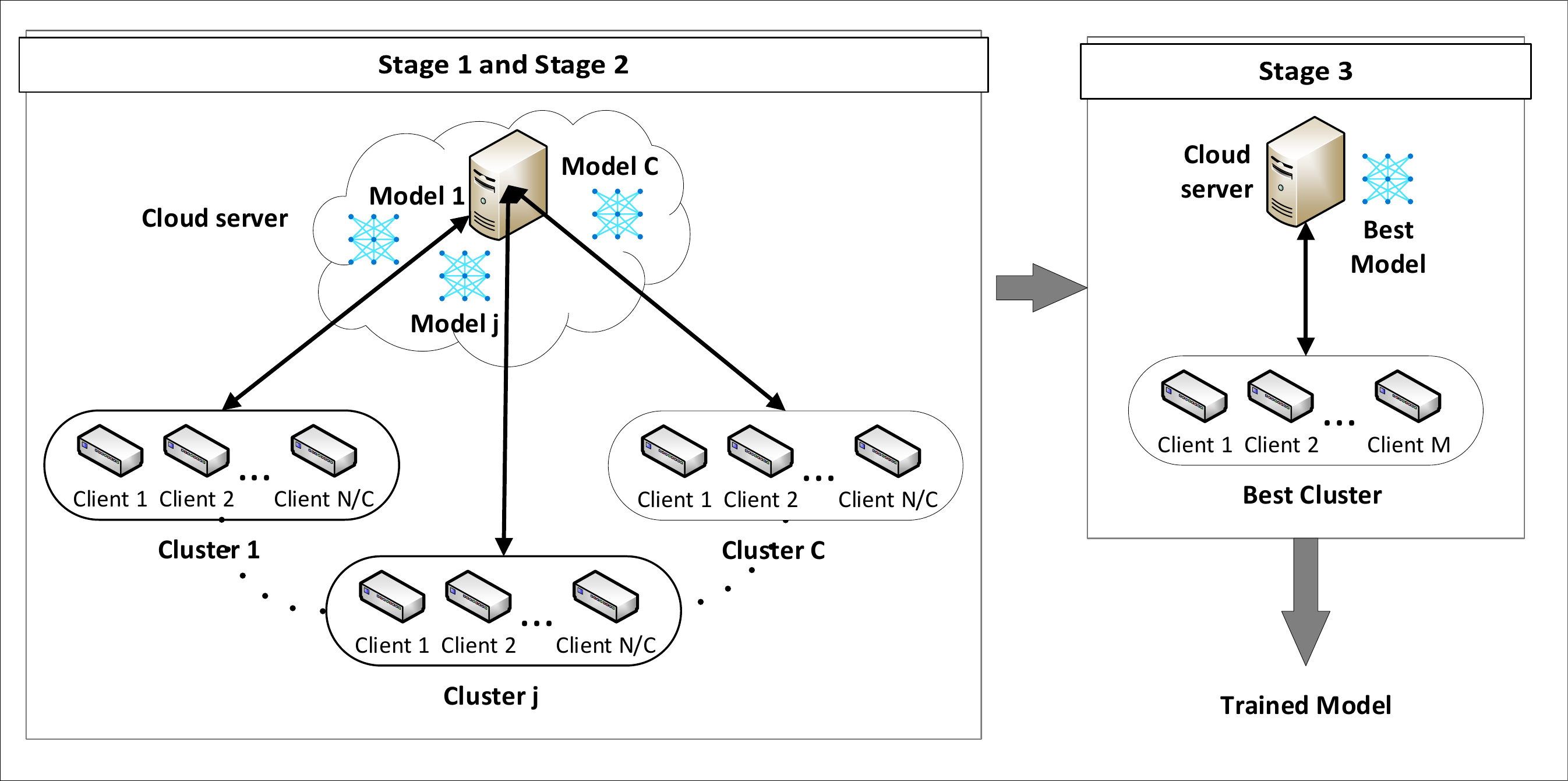}
\caption{An illustration of the system model. The cloud server running the proposed algorithm communicates with the local servers in different clusters to train a number of global models in stage 2 then the best model is trained in stage 3.}
\label{fig:application}
\end{figure}

We assume one server, $N$ clients, and $C$ clusters such that there are $N/C$ clients per cluster except for the last cluster, which may have fewer clients as shown in Fig. \ref{fig:application}. We also assume a training budget of $R$ rounds. There are three stages for the system to find and train the best model. In the first stage, the server randomly assigns clients to $C$ clusters and selects a random model for each cluster. In other words, the server assigns the same model to all clients of the same cluster. Additionally, the server has a small unlabeled dataset used for testing models during the training process to determine the performance of models and thus determining the best and worst-performing models. The use of an unlabeled dataset rather than a labeled one ensures the privacy of data on the server. In the second stage, the system runs $C-1$ iterations to explore models. In each iteration, clients in each cluster are engaged with the server for a number of communication rounds to train the global model of that specific cluster. By the end of each iteration, two actions take place. First, some clients (depending on the value of $X$ as explained later) are expelled from each cluster due to a poisoned or poor dataset. Second, the best and worst-performing models (i.e., clusters) are determined, and the worst cluster is removed, and its clients are exploited and assigned to the best performing cluster. By the end of the last iteration, only one cluster remains with $M \leq N$ clients as shown in Fig. \ref{fig:application}. In the third stage, clients in the remaining cluster engage with the server for a number of iterations to train the remaining global model.

\setlength{\textfloatsep}{4pt}

\begin{algorithm}[htbp]
\footnotesize
\caption{Proposed heuristic}
\label{alg_proposed}
\begin{algorithmic}[1]
\STATEx \textbf{Input}: $N$: number of clients, $C$: number of clusters, $R$: number of communication rounds, $E$: number of epochs, $R_c$: number of communication rounds per cluster, $X$: percentage of expelled clients per cluster per phase
\STATEx \textbf{Output}: Trained global model

\STATEx
\begin{center}
\textbf{\textit{// Server initialization}}
\end{center}
\FOR {$c$ = 1 to $C$}
\STATE Pick random model $M_c$ for cluster $c$
\STATE Initialize the global model $M_c$
\STATE Pick $N_c \leq N/C$ clients for cluster $c$
\ENDFOR
\STATE Set $E_{stage2}$ = ($R$ / 2) x $E$ 
\STATE Set $E_c$ = $E_{stage2}$ / ($C$ - 1) / $R_c$
\STATEx
\begin{center}
\textbf{\textit{// Find the best model in terms of predicted labels}}
\end{center}
\WHILE {$C$ $>$ 1}
\FOR {$c$ = 1 to $C$}
\FOR {$r=1$ to $R_c$}
\STATE Server send global model $M_c$ to all clients in $c$
\STATE Clients train global model on local dataset with $E_c$ epochs and return updated model
\STATE Server average aggregated model parameters from clients
\ENDFOR
\STATE Predict labels of the unlabeled dataset using the global model on the server
\STATE Predict labels of the unlabeled dataset for every returned clients' model on the server
\STATE Compute cosine similarity for every returned clients' models
\STATE Exclude $X\%$ of clients with the lowest cosine similarity from cluster $c$
\ENDFOR
\STATE Compute the average of predicted labels using global models of all clusters
\STATE Compute cosine similarity for predicted labels of every cluster against the average predicted labels of all clusters
\STATE Find $c_h$ the cluster with the highest cosine similarity
\STATE Find $c_l$ the cluster with the lowest cosine similarity
\STATE Move all clients from $c_l$ to $c_h$
\STATE Delete cluster $c_l$
\STATE Set $C=C-1$
\ENDWHILE

\STATEx
\begin{center}
\textbf{\textit{// Train the last remaining cluster}}
\end{center}
\STATE Set c = last remaining cluster number
\FOR {$r=1$ to $R/2$}
\STATE Server send global model $M_c$ to all clients in $c$
\STATE Clients train global model on local dataset with $E$ epochs and return updated model
\STATE Server average aggregated model parameters from clients
\ENDFOR
\STATE Return the trained global model
\end{algorithmic}
\end{algorithm}

\vspace{-5mm}

\section{Proposed Algorithm}
\label{proposed_algorithm}

The proposed algorithm consists of three stages. In the first stage (Algorithm~\ref{alg_proposed}, lines 1 through 7), $C$ clusters are formed and each cluster hosts about $N/C$ clients and uses a model selected randomly. We created a pool of random models to select from. Each model in the pool is created with three parameters: number of convolutional layers (1 or 2), filters (64, 128, 196, or 256), and kernel size (3x3, 3x5, 5x3, or 5x5). There is an input layer with 28x28 size. Also, a max-pool layer is used after every convolutional layer. The last 3 layers are one flatten layer and two dense layers.

In the second stage (Algorithm~\ref{alg_proposed}, lines 8 through 27), models are explored, and the best performing model (i.e., cluster) is selected in a repetitive process with $C-1$ iterations. In each iteration, all clients in all clusters are trained in parallel for $R_c$ communication rounds, then $X\%$ clients are expelled from each cluster due to low performance or poisoned data. Next, the cluster with the most deficient performance is deleted, and its clients are exploited by joining the highest performing cluster. This process continues until only one cluster is left. In the third stage (Algorithm~\ref{alg_proposed}, lines 28 through 34), clients in the remaining cluster are trained for $R/2$ communication rounds, and the trained model is returned.

We can summarize the proposed algorithm (Algorithm \ref{alg_proposed}) as follows:

\begin{itemize}
    \item Lines 1 through 5: initialize a random model and select $N/C$ clients randomly for each of the $C$ clusters.
    \item Lines 6 through 7: set training parameters per cluster
    \item Lines 8 through 27: find the best model by running $C-1$ phases. In each phase, clients of all available clusters are trained in parallel. Clients are evaluated in every cluster based on the cosine similarity of predicted labels of the global model against predicted labels of every client's model using an unlabeled dataset in the server. Then, $X\%$ clients with the lowest cosine similarity are expelled from every cluster. The low performance of expelled clients is either related to poor data or poisoned data. Finally, the clusters are evaluated based on their average predicted labels, and the cluster with the lowest cosine similarity is deleted with its clients moved to the cluster with the highest cosine similarity.
    \item Lines 28 through 34: train the remaining clusters and return the trained model.
\end{itemize}

\vspace{-3mm}


\section{Experimental Settings}
\label{experimental_settings}

\subsection{Dataset}
\label{dataset}
We use the Federated Extended MNIST, \textbf{FEMNIST} \cite{caldas2018leaf}, for classifications tasks. The FEMNIST is used for both letters and digits (A-Z, a-z, and 0-9), and it has 244,154 images for training and 61500 for testing. We use this dataset under non-i.i.d data distribution, where the FEMNIST dataset is first split into 62 partitions (number of labels). Then each of the 900 users is assigned batches of two classes only.

For adversarial versions: we applied Fast Gradient Sign Method  \textbf{(FGSM)} proposed by Goodfellow \cite{goodfellow2014explaining}. FGSM is widely used to produce adversarial examples. The original input is manipulated by adding or subtracting a small error of $\epsilon$ to each data sample. $\epsilon$ is a small number controlling the size of the adversarial attack to be effective. Any addition or subtraction of the $\epsilon$ depends on the gradient sign for any given input that is either positive or negative. Adding errors in the gradient direction means that classification is intentionally altered so that the model classification fails.

We divided the training dataset over $N$ clients in all experiments and used a random unlabeled number of records from the test dataset on the server, which is used for evaluating clients' local models and global cluster models.

\subsection{Experiments}
\label{experiments}

To evaluate the performance of the proposed algorithm, we compared the results of the proposed algorithm against the random FL algorithm (baseline algorithm), which is also used as the baseline algorithm in \cite{mohammed_budgeted_2020}. To ensure a fair comparison, we use the same number of epochs for both algorithms. We set $R$, the number of communication rounds to 32, and the number of epochs $E$ to 8 for all experiments of the baseline algorithm, so every client uses a total of 256 (8 x 32) epochs. We used less than or the equal number of epochs ($\leq$ 256) with all experiments of the proposed algorithm.

For example, assume that we have 8 clusters. Since there are 256 total epochs available for every client in the baseline algorithm, we use half of it or less in the second stage of the proposed algorithm and use the other half in the third stage of the proposed algorithm. To do that, we first compute the total number of epochs available for the second stage of the proposed algorithm as shown in equation \eqref{eq1}, which is 128.

\begin{equation}
E_{stage2}=\frac{R}{2}E
\label{eq1}
\end{equation}

Since we have $C-1$ iterations, we need to compute the number of epochs per client in each iteration, and the total number of epochs per client for the $C-1$ iterations must be $\leq$ 128. Assuming that $R_c$, the number of communication rounds per cluster is 4, we compute the total number of epochs used by every client per iteration in the second stage of the proposed algorithm as shown in equation \eqref{eq2}. This number is 4.57, so we round down the number to 4 epochs. Consequently, the second stage of the proposed algorithm is using only 112 epochs (4 epochs times 4 rounds times 7 iterations) and not 128.

\begin{equation}
E_c=\frac{E_{stage2}}{(C-1) R_c}
\label{eq2}
\end{equation}

In the third stage of the proposed algorithm, we use $R/2$ times $E$ epochs, which is 128. Thus, the proposed algorithm is actually using less number of epochs, 240 in this example, compared to the baseline algorithm. In other words, the proposed algorithm consumes fewer computing resources compared to the baseline algorithm. In general, the proposed algorithm uses $\leq$ epochs compared to the baseline algorithm.

We conducted 208 total experiments using the parameters shown in Table \ref{table_Sim_parameters}. We run 16 experiments with the baseline algorithm using all combinations of $N$ and $P_{perc}$. Then we run 192 experiments with the proposed algorithm using all combinations of $N$, $C$, $P_{perc}$, and $X_{perc}$.

Running those experiments on a single computer takes months. Thus, we utilized tens of nodes in the Holland Computing Center at the University of Nebraska \cite{noauthor_holland_nodate}. We run all of the 16 experiments of the baseline algorithm in parallel as a single batch, then divided the 192 experiments of the proposed algorithm into 8 batches, each having 24 experiments that run in parallel.

\begin{table}[htbp]
 \footnotesize
 \centering
 \caption{Simulation Parameters.}
 \label{table_Sim_parameters}
 \begin{tabular}{c|c|c}
 \hline
 \thead{\textbf{\textit{Sym.}}} & \thead{\textbf{\textit{Parameter}}} & \thead{\textbf{\textit{Value(s)}}}
 \\ \hline
 $N$ & No. of clients & 
 (100, 200, 400, 900)
 \\ \hline
 $C$ & No. of clusters &
 (4, 8, 16, and 32)
 \\ \hline
 $R$ & Communication rounds & 32
 \\ \hline
 $E$ & Epochs & 8
 \\ \hline
 $B$ & Batches & 32
 \\ \hline
 $R_c$ & Rounds per cluster & 4
 \\ \hline
 $P$ & Percentage of poisoned dataset & (0, 10, 20, and 40)
 \\ \hline
 $X$ & Percentage of expelled clients & (0, 10, and 20) \\
 & per cluster &
 \\ \hline
 \end{tabular}
\end{table}

\begin{figure}[htbp]
\centering
\begin{subfigure}[b]{0.35\textwidth}
\includegraphics[width=\textwidth]{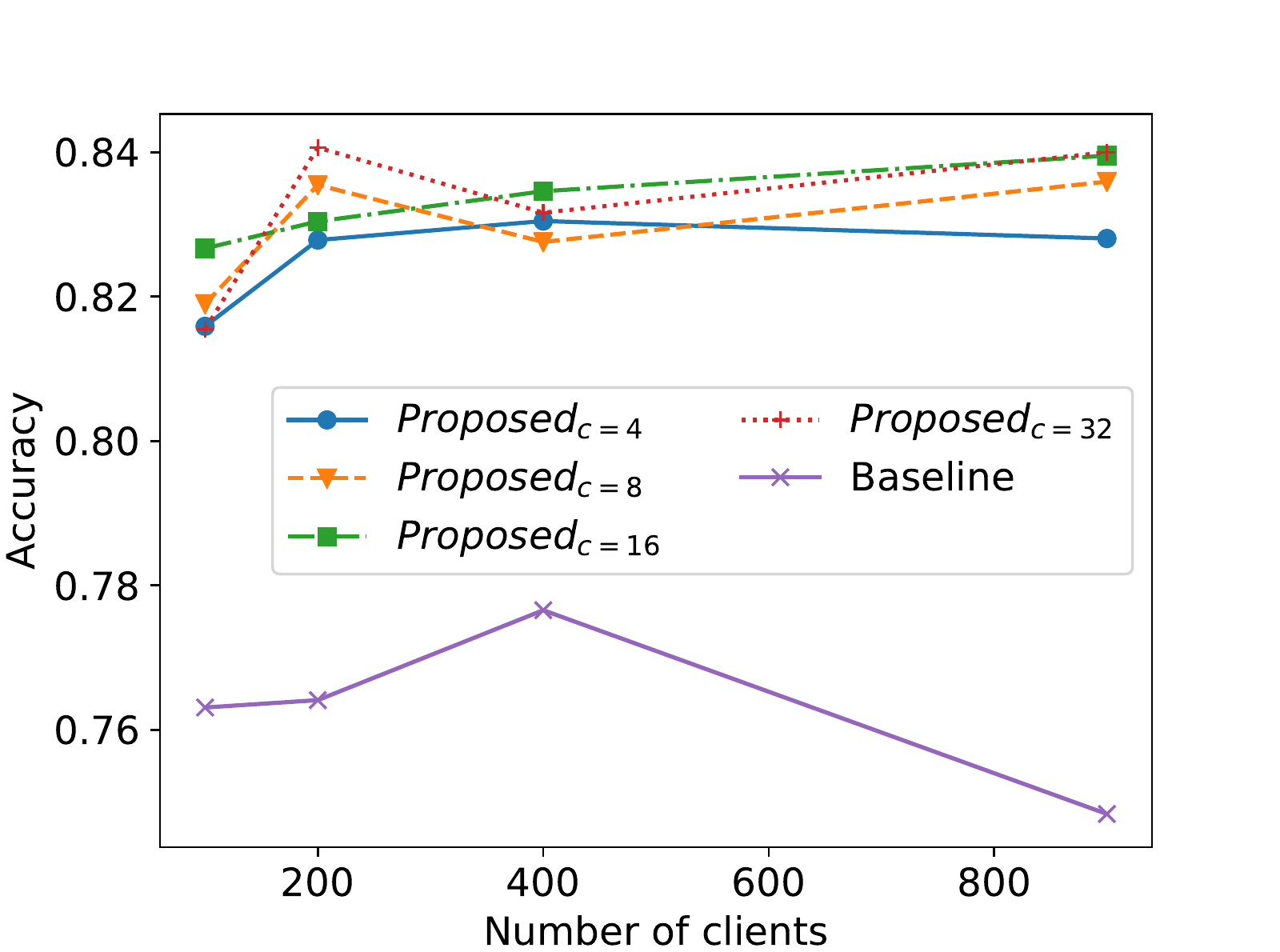}
\caption{Accuracy of the proposed v.s. the baseline algorithms with $P=0\%$ and $X=0\%$.}
\label{fig:clustering}
\end{subfigure}
\begin{subfigure}[b]{0.35\textwidth}
\includegraphics[width=\textwidth]{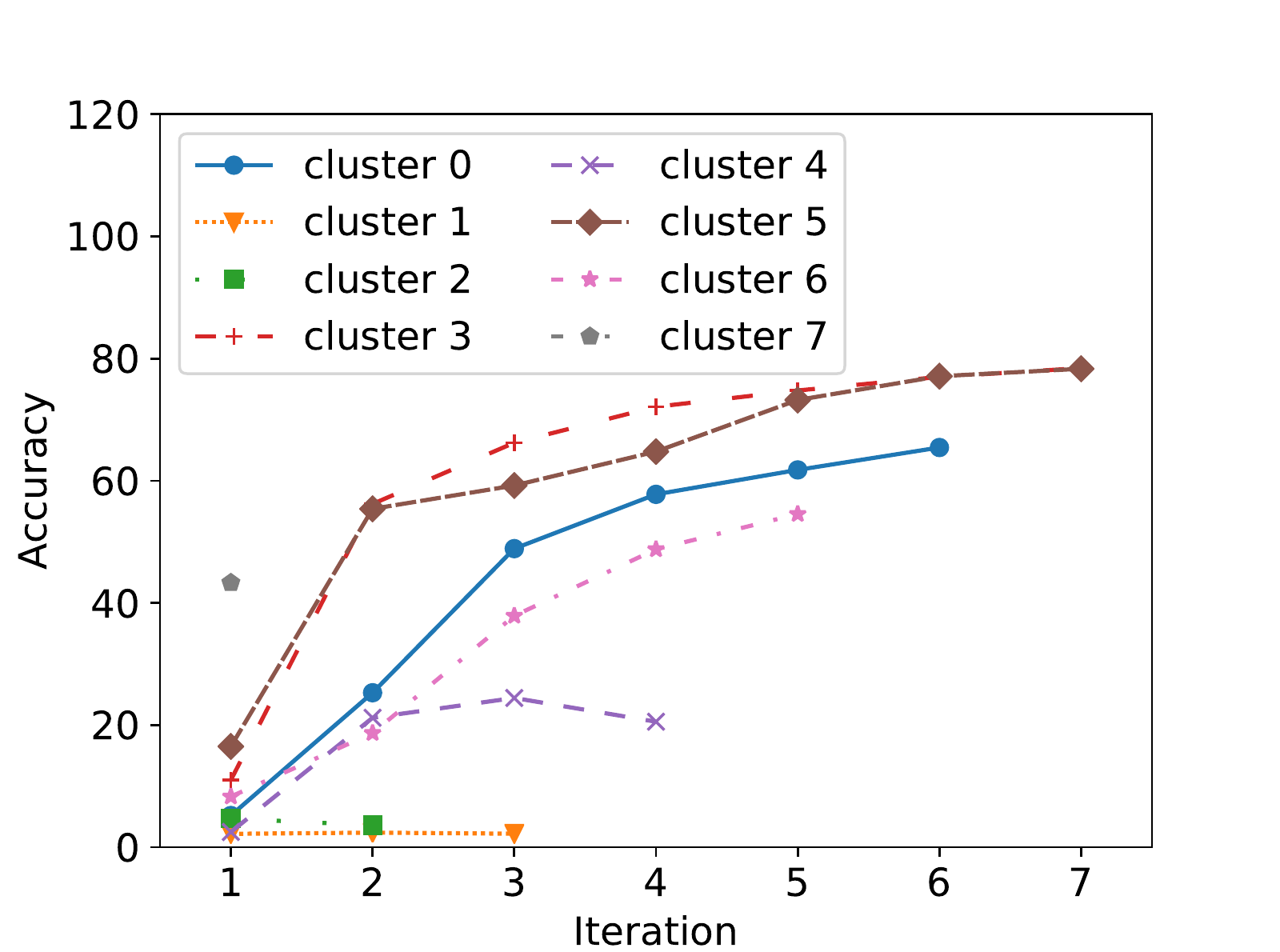}
\caption{Accuracy of models (i.e clusters) used in the proposed algorithm with $N=400$, $P=40\%$, and $X=20\%$.}
\label{fig:iterations}
\end{subfigure}
\begin{subfigure}[b]{0.35\textwidth}
\includegraphics[width=\textwidth]{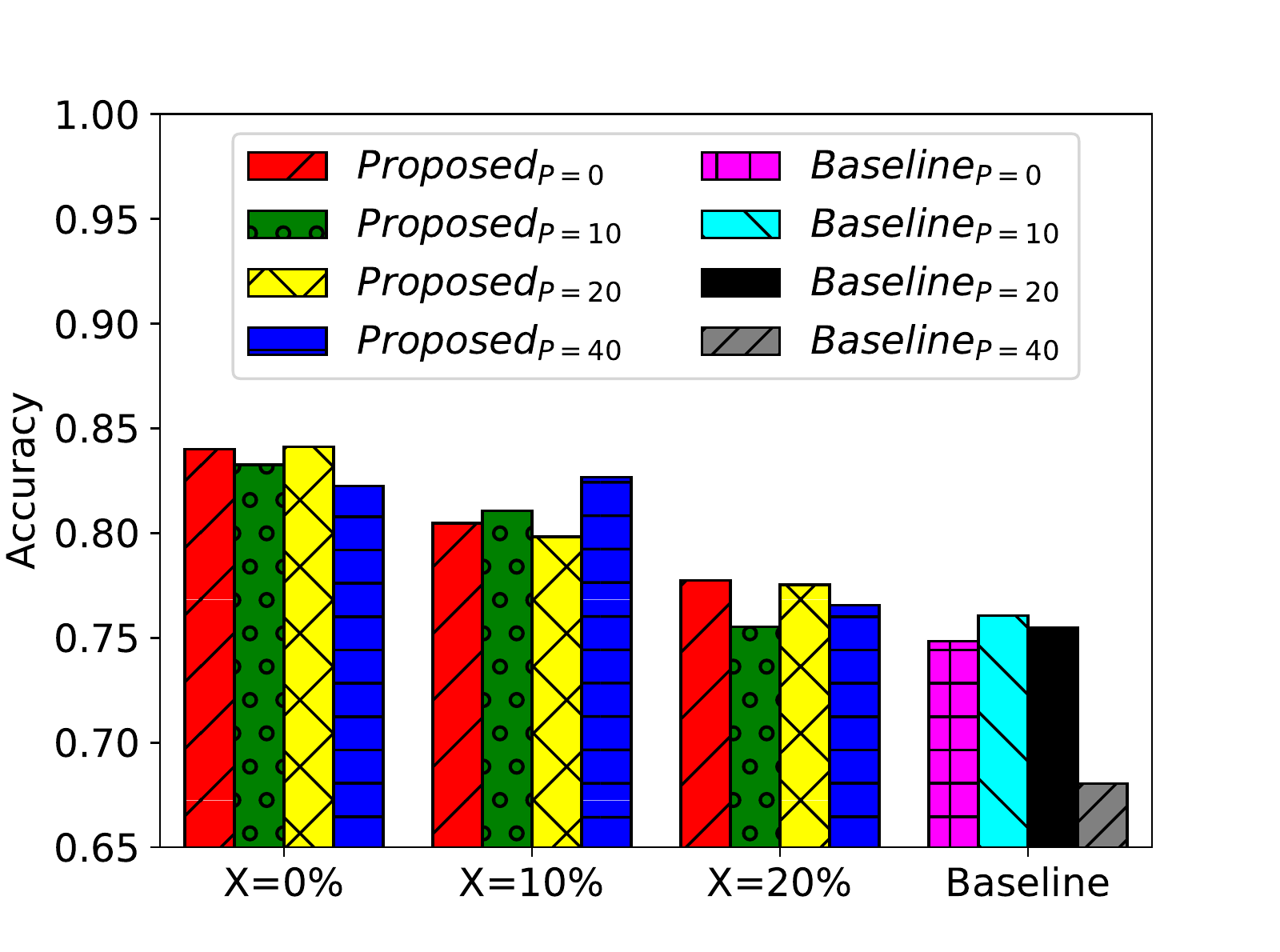}
\caption{Accuracy of the proposed v.s. the baseline algorithms with $C=32$ and $N=900$.}
\label{fig:poisoned}
\end{subfigure}
\begin{subfigure}[b]{0.35\textwidth}
\includegraphics[width=\textwidth]{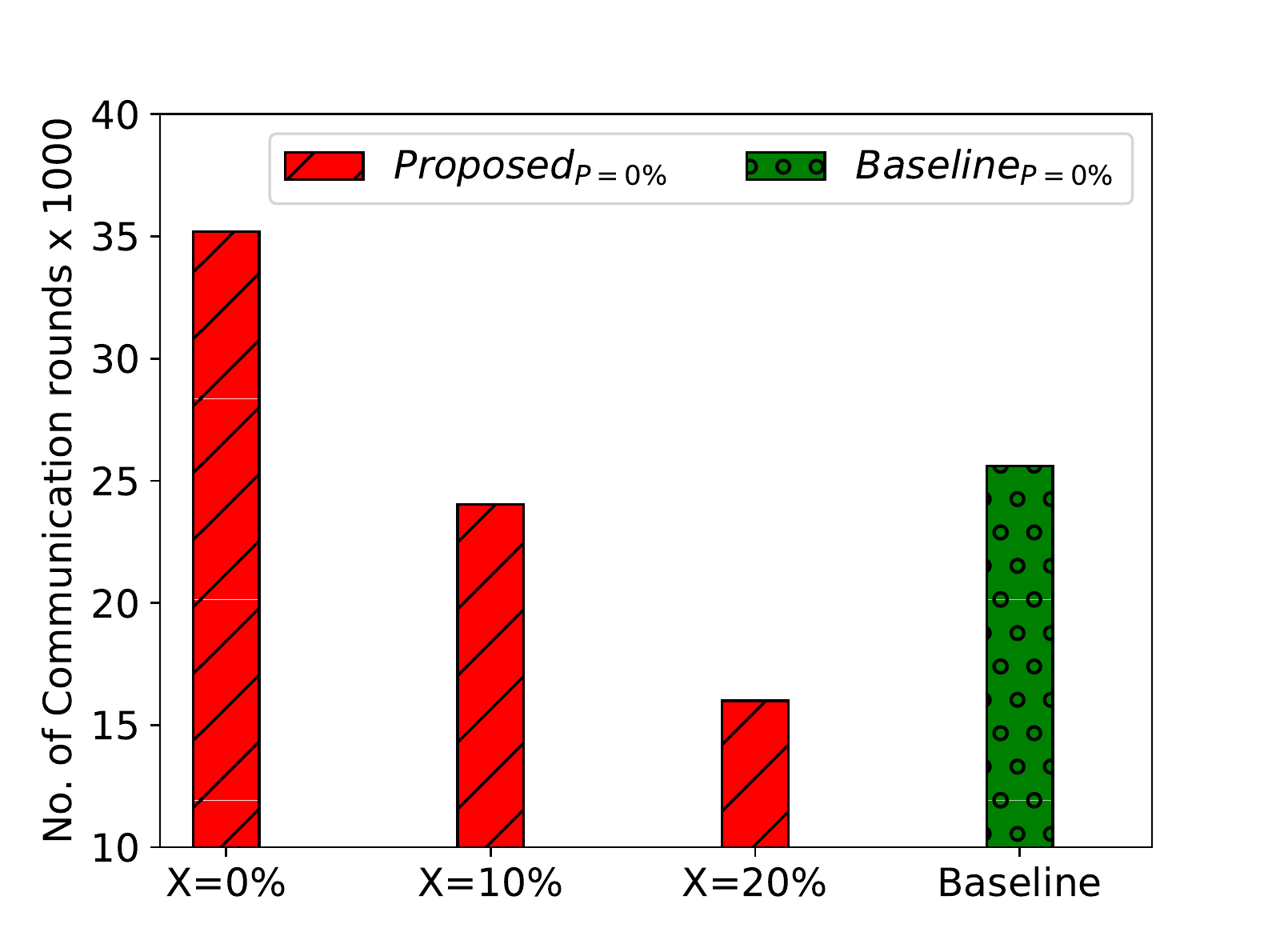}
\caption{Communication cost of the proposed v.s. the baseline algorithms with $N=400$, $P=0\%$, and $C=8$.}
\label{fig:communication}
\end{subfigure}
\caption{Comparison of the proposed FL algorithm against the baseline.}
\label{fig:N}
\end{figure}

\vspace{-5mm}

\section{Results Discussion}
\label{result_discussion}

Since we cannot present all 208 experiments, we fixed some parameters and show the results for changing the other parameters.

\subsection{Performance of Clustering}
\label{perf:clustering}

We measured the performance of both algorithms in terms of accuracy using a non-poisoned dataset ($P=0\%$) and disabled expelling in the proposed algorithm ($X=0\%$). Results are illustrated in Fig. \ref{fig:clustering} (summary of 20 experiments), which shows the superior performance of the proposed algorithm over the baseline algorithm using different numbers of clusters and clients. These results support the claim that the second stage of the proposed algorithm selects a better model than the baseline algorithm and thus results in better (i.e. higher) accuracy while using the same number of epochs. However, this gain in performance comes at the expense of communication cost, which can be reduced when clients are expelled (i.e., $X\geq0$) as discussed in subsection \ref{perf:expelling}. Fig. \ref{fig:iterations} shows the second stage of the proposed algorithm in action. It shows the accuracy of each model (i.e., cluster) over iterations, starting with 8 models at iteration 1 and ending with one model (best performing model in cluster 5) at iteration 7.

\subsection{Performance with Poisoned Dataset}
\label{perf:poisoned}

To study the effect of the poisoned dataset on the performance of the two algorithms, we run both algorithms using different percentages of the poisoned dataset as shown in Fig. \ref{fig:poisoned}. In this figure, we can see that when the poisoned percentage of the dataset is higher, especially when $P=40\%$, the performance of the baseline algorithm in terms of accuracy is severely impacted. On the other hand, the proposed algorithm is barely impacted due to clustering (i.e. better performing model) and expelling of clients with the poisoned dataset.

\begin{table}[htbp]
 \footnotesize
 \centering
 \caption{Efficiency of expelling clients with $N=400$, $P=40\%$, $X=20\%$, and $C=8$.}
 \label{table_neg_pos}
 \begin{tabular}{c|c|c|c|c|c}
 \hline
 \thead{\textbf{\textit{Iteration}}} & \thead{\textbf{\textit{True}} \\ \textbf{\textit{Positive}}} &
 \thead{\textbf{\textit{False}} \\ \textbf{\textit{Positive}}} &
 \thead{\textbf{\textit{True}} \\ \textbf{\textit{Negative}}} &
 \thead{\textbf{\textit{False}} \\ \textbf{\textit{Negative}}} &
 \thead{\textbf{\textit{Total}} \\ \textbf{\textit{Nodes}}}
 
 \\ \hline
 1 & 24 & 56 & 184 & 136 & 400
 \\ \hline
 2 & 31 & 33 & 151 & 105 & 320
 \\ \hline
 3 & 37 & 11 & 140 & 68 & 256
 \\ \hline
 4 & 31 & 9 & 131 & 37 & 208
 \\ \hline
 5 & 20 & 12 & 119 & 17 & 168
 \\ \hline
 6 & 16 & 10 & 109 & 1 & 136
 \\ \hline
 7 & 1 & 20 & 89 & 0 & 110
 \\ \hline
 \end{tabular}
\end{table}

\vspace{-5mm}

\begin{figure}[htbp]
\centering
\includegraphics[width=2in]{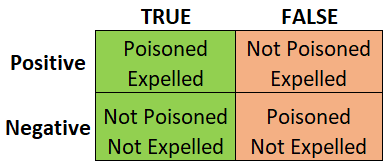}
\caption{Meaning of Negative, Positive, True, and False in Table \ref{table_neg_pos}}
\label{fig_neg_pos}
\end{figure}

\vspace{-5mm}

\subsection{Performance with Expelling Clients}
\label{perf:expelling}

Expelling clients with poisoned or poor dataset prevent the deterioration of the performance of the proposed algorithm as indicated in Fig. \ref{fig:poisoned} compared to the baseline algorithm. However, expelling a client with a good dataset can reduce the performance of the proposed algorithm. To study the expelling process in more depth, we track the number and status of expelled clients after each iteration and create a confusion matrix as shown in Table \ref{table_neg_pos}. Fig. \ref{fig_neg_pos} explains the meaning of the four main columns in Table \ref{table_neg_pos}. For example, a True Positive number represents the number of clients being expelled by the proposed algorithm that has a poisoned dataset. A False Positive is not always a bad indication because the expelled clients may have a poor dataset. On the other hand, a False Negative always negatively impact the performance of the proposed algorithm. We start with 400 nodes in the experiment in Table \ref{table_neg_pos} with a total of 160 poisoned nodes (40\%). In the first iteration, the number of False Positive is higher than the number of True Positive, which is expected since the proposed algorithm is just starting and need more training. Also, the number of False Negative is high because the proposed algorithm is defined to expel only 20\% after each iteration and thus cannot eliminate all poisoned clients in one iteration. In the second iteration and on, the number of True Positive becomes close or higher than the number of False Positive, which proves that the proposed algorithm is working as expected and detects poisoned clients more accurately. After the last iteration, we have only 110 clients left out of the 400, 0 False Negative, and all of the 160 clients with poisoned datasets are expelled successfully (i.e., 100\%). Out of the 290 expelled clients, there are 130 clients with a clean dataset, and those have poor datasets compared with the remaining clients.

The proposed algorithm uses only 110 clients out of 400 in the third stage and still gets better results compared with the baseline algorithm in terms of accuracy. This big reduction in the number of utilized clients reduces the communication cost tremendously, as illustrated in Fig. \ref{fig:communication}.

\subsection{Lessons Learned} 

The key lessons learned from the experiments conducted in this work can be summarized as follows.

\begin{itemize}
    \item The proposed algorithm can select a better model compared to the baseline algorithm due to using exploration and exploitation and thus results in better accuracy while using the same number of epochs.
    \item The proposed algorithm is better suited to cope with poisoned data, compared to the conventional FL algorithm, by expelling clients with the poisoned dataset. 
    \item The proposed algorithm can accurately identify the clients with poisoned or poor data without affecting the overall performance of the final model.
    \item The expelling process not only expels clients with poisoned and poor dataset but also reduce communication cost.
\end{itemize}

\section{Conclusions and Future Work}
\label{conclusion_future_work}

In this paper, we presented an algorithm that clusters clients in the first stage into a number of clusters, each with a random model. In the second stage, the proposed algorithm explores those models (i.e., clusters) by training them in a number of iterations and reduces the number of clusters by one after each iteration to find the best performing model (i.e., cluster). Also, in each iteration, the proposed algorithm expels some clients that have poisoned or poor datasets while surviving clients are exploited in the next iterations. Then, in the third stage, the proposed algorithm continues the training process with the one remaining cluster, representing the selected model (best performing model) and returning the trained selected global model.
The proposed algorithm is compared with a baseline algorithm, which is the random FL. Results show that the proposed algorithm is performing better in terms of accuracy and number of communication rounds when configured to expel clients compared with the baseline algorithm.

In the future, we plan to solve the selection of clients as an optimization problem to maximize the accuracy and minimize communication rounds given a fixed percentage of the poisoned dataset.

\section*{Acknowledgment}
This publication was made possible by NPRP grant \# [13S-0206-200273] from the Qatar National Research Fund (a member of Qatar Foundation). The statements made herein are solely the responsibility of the authors.
\ifCLASSOPTIONcaptionsoff
 \newpage
\fi

\bibliographystyle{IEEEtran}
\bibliography{biblo}

\end{document}